\documentclass[journal=jpcafh,manuscript=article]{achemso}

\usepackage[version=4]{mhchem} 
\usepackage{graphicx}
\usepackage{color}
\usepackage{dcolumn}
\usepackage{multirow}
\usepackage[normalem]{ulem}
\usepackage{cmbright}
\usepackage[dvipsnames]{xcolor}
\SectionNumbersOn

\newcommand{\atmo}{\textsc{AtmoSpec}}
\newcommand{\aiida}{\textsc{AiiDA}}
\newcommand{\aiidalab}{\textsc{AiiDAlab}}

\author{Daniel Hollas}
\email{daniel.hollas@bristol.ac.uk}
\author{Basile F. E. Curchod}
\email{basile.curchod@bristol.ac.uk}
\affiliation[University of Bristol]
{Centre for Computational Chemistry, School of Chemistry, University of Bristol, Cantock’s Close, Bristol BS8 1TS, United Kingdom}

\title{\atmo{} -- A Tool to Calculate Photoabsorption Cross-Sections for Atmospheric Volatile Organic Compounds}

\begin{document}

\begin{tocentry}
    \includegraphics[width=1.0\textwidth]{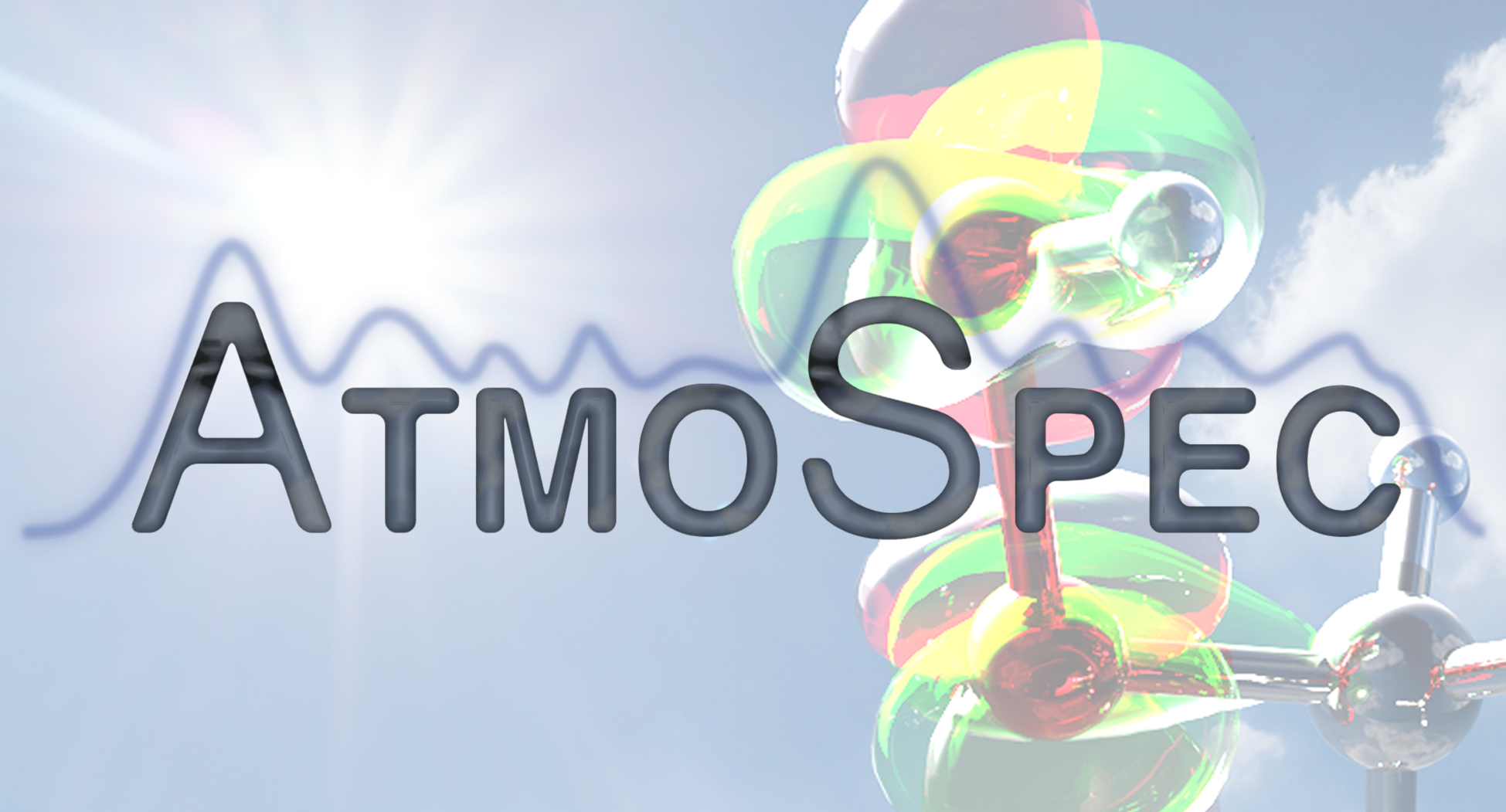}

\end{tocentry}

\begin{abstract}
Characterizing the photolysis processes undergone by transient volatile organic compounds in the troposphere requires the knowledge of their photoabsorption cross-section -- quantities often challenging to determine experimentally, particularly due to the reactivity of these molecules. We present a computational tool coined \atmo{}, which can predict a quantitative photoabsorption cross-section for volatile organic compounds by using computational photochemistry. The user enters the molecule of interest as a SMILES code and, after selecting a level of theory for the electronic structure (and waiting for the calculations to take place), is presented with a photoabsorption cross-section for the low-energy conformers and an estimate of the photolysis rate coefficient for different standardized actinic fluxes. More specifically, \atmo{} is an automated workflow for the nuclear ensemble approach, an efficient technique to approximate the absolute intensities and excitation wavelengths of a photoabsorption cross-section for a molecule in the gas phase. This work provides background information on the nuclear ensemble approach, a guided example of a typical \atmo{} calculation, details about the architecture of the code, and the current limitations and future developments of this tool.
\end{abstract}

\section{Introduction}

Understanding and predicting the photochemical reactivities and lifetimes of atmospheric volatile organic compounds (VOCs) upon sunlight absorption is of prime importance in informing chemical models used to simulate the molecular composition of our atmosphere.\cite{egusphere-2024-1316,curchodorrewing2024} However, the potential for photolysis of a large number of VOCs, in particular \textit{transient} VOCs, remains unknown to date due to their short lifetime and instability, preventing their isolation or synthesis and rendering spectroscopic measurements highly challenging. 

The photolysis of a compound \ce{A} following sunlight absorption, \ce{A ->[h\nu] B + C}, is characterized by the photolysis rate $-\frac{d[A]}{dt}=J[A]$, where $J$ is the photolysis rate coefficient. This photolysis rate coefficient is defined as

\begin{equation}
J=\int_{\lambda_\text{min}}^{\lambda_\text{max}} \phi(\lambda)\sigma(\lambda) F(\lambda)d\lambda.
\label{eq:photolysis}
\end{equation}

Equation~\eqref{eq:photolysis} provides the story of a photolysis process: the actinic flux ($F(\lambda)$) informs on the number of photons coming from a source (e.g., the sun) at a given wavelength per second and per unit area, and the likelihood of molecule \ce{A} to absorb one of these photons is given by its photoabsorption cross-section $\sigma(\lambda)$. If the molecule has absorbed light at wavelength $\lambda$, the quantum yield $\phi(\lambda)$ gives the probability that the photoexcited molecule will undergo the photolysis process described above and form \ce{B + C}. Integrating the product $\phi(\lambda)\sigma(\lambda) F(\lambda)$ over a range of wavelength characteristic of the actinic region (280 - 400 nm) leads to the overall photolysis rate coefficient $J$. (We note at this point that $J$ often depends on the temperature, pressure, and the solar zenith angle.)

Equation~\eqref{eq:photolysis} makes it clear that a key quantity to consider when studying the potential for photolysis of a VOC is its photoabsorption cross-section $\sigma(\lambda)$. Photolysis data for a limited number of key VOCs can be found in databases like MPI-Mainz UV/Vis Spectral Atlas,\cite{keller2013mpi} IUPAC,\cite{ACPcopernicusWebsite} NASA (JPL),\cite{burkholder2020chemical} or atmospheric models such as TUV.\cite{TUVwebsite} Information about the potential photolysis of VOCs is crucial to inform atmospheric models like the Master Chemical Mechanism used to simulate the chemical composition of the atmosphere.\cite{jenkin1997tropospheric,saunders2003,jenkin2003,MCMwebsite} As stated above, obtaining such quantities experimentally is often hampered by the instability of the molecule (or the difficulty in isolating it), and structure-activity relationships (SARs) -- regularly used for ground-state properties of VOCs -- often run out of steam when extended to photochemical processes, notably due to the possible delocalized nature of molecular excited states and the multichromophoric nature of some VOCs.\cite{prlj2021calculating}  

Recently, we proposed to use computational photochemistry to obtain an estimate of the photoabsorption cross-section of transient VOCs.\cite{prlj2020theoretical,prlj2021calculating} The computational strategy, based on the nuclear ensemble approach (NEA, see Sec.~\ref{sec:NEA} below for details),\cite{crespo2014spectrum,schinkebook} offers an adequate description of the absolute cross-section and the position of the maxima in the photoabsorption cross-section for VOCs. While this technique does not describe vibronic progressions,\cite{crespo2014spectrum,crespo2018recent,mcgillen2017criegee} the width of each band is captured qualitatively well, which is a prime factor when considering the use of calculated photoabsorption cross-sections in Equation~\eqref{eq:photolysis}.

Hence, the NEA could act as a doorway to predict the photoabsorption cross-section of transient VOCs and, hopefully, determine more robust photochemical SARs rules for compounds of the same family. However, the practical execution of the NEA workflow is rather tedious (yet not complex), which hinders its adoption outside of the computational photochemistry community. Here, we introduce \atmo{} -- an automated workflow and graphical interface to predict the photoabsorption cross-sections of VOCs based on the NEA and requiring only a limited number of operations from the user, making the theoretical prediction of cross-sections more accessible to researchers outside of the computational photochemistry community.

In the following, we briefly summarize the main conceptual ideas of the NEA, the different steps required for a typical calculation, and the limitations of the methods (Sec.~\ref{sec:NEA}).  We then walk the reader through an example calculation in \atmo{} to determine the photoabsorption cross-section of glycolaldehyde (Sec.~\ref{sec:demonstration}). We present the design and architecture of \atmo{} and the details of the automatized NEA workflow implemented using the \aiida{} infrastructure (Sec.~\ref{sec:architecture}). We finally stress the current limitations of \atmo{} and highlight its future developments (Sec.~\ref{sec:caveats}). 

\section{A brief survey of the nuclear ensemble approach}
\label{sec:NEA}

Different theoretical strategies can be used to predict the photoabsorption of gas-phase molecules like VOCs (see Ref.~\citenum{prlj2021calculating} for an overview), but here we focus exclusively on the NEA which represents a good tradeoff between computational efficiency and qualitatively robust results. Crucially, unlike other more complicated approaches, the NEA workflow is amenable to automation. The NEA, which is a numerical realization of the reflection principle,\cite{schinkebook} proposes to project the ground-state nuclear density onto the excited electronic state(s) of interest.\cite{crespo2014spectrum} In practice, the NEA consists of (\textit{i}) approximating the ground-state probability density of the molecule of interest, (\textit{ii}) sampling from this distribution a set of $N_p$ molecular geometries ($N_p \approx 500-10000$), and (\textit{iii}) calculating, for each molecular geometry $j$ ($\mathbf{R}_j$, where $\mathbf{R}$ is a collective variable for all the nuclear coordinates of the molecule of interest), the corresponding vertical excitation energies ($\Delta E_{0I}(\mathbf{R}_j)$) and transition dipole moments ($\boldsymbol{\mu}_{0I}(\mathbf{R}_j)$) to a subset of $N_s$ excited electronic states of interest (labeled by $I$). For each geometry, each transition energy is broadened by a Gaussian (or Lorentzian) with a set width $\delta$. The photoabsorption cross-section is obtained by averaging over all the $N_s$ broadened electronic transitions for all the N$_p$ geometries. Hence, the photoabsorption cross-section within the NEA for a given conformer is obtained as 
\begin{equation}
     \sigma(E) = \frac{\pi}{3\hbar \epsilon_0 c} \frac{1}{N_p \delta \sqrt{2\pi}} \sum_{j=1}^{N_p} \sum_{I=1}^{N_s} \Delta E_{0I}(\mathbf{R}_j) |\boldsymbol{\mu}_{0I}(\mathbf{R}_j)|^2 \exp{\left(-\frac{\left(E-\Delta E_{0I}(\mathbf{R}_j)\right)^2}{2\delta^2}\right)}   \, ,
\label{eq:nea}
\end{equation}

with $E$ being the photon energy. 

If the molecule of interest exists as multiple conformers at a given temperature, the protocol described above should be repeated for each conformer, and the total photoabsorption cross-section is obtained by summing the Boltzmann-weighted photoabsorption cross-section for each conformer.

The central idea of the NEA -- to project a ground-state probability density vertically onto excited electronic state(s) -- is justified for dissociative excited electronic states for which the corresponding photoabsorption cross-section does not exhibit vibronic progression.\cite{schinkebook} In other words, the NEA cannot reproduce vibronic progressions as it does not consider the vibrational states of the excited electronic states.\cite{crespo2018recent,prlj2021calculating} For dissociative excited states ($n\sigma^\ast$, $\pi\sigma^\ast$, $\sigma\sigma^\ast$), which are of central interest for photolysis processes related to atmospheric photochemistry, the NEA can provide an accurate description of the photoabsorption cross-sections (absolute intensities and excitation wavelengths), given that an adequate electronic-structure method was employed for excitation energies and transition dipole moments -- see Refs.\citenum{mcgillen2017criegee,srvsevn2018uv,prlj2021calculating,mccoy2021simple,clarke2022photochemistry,prlj2023deciphering} for examples. However, the NEA was shown to provide a qualitatively correct description of the transition bands for photoabsorption cross-sections involving bound excited electronic states (like $n\pi^\ast$), capturing the proper envelope of each band as well as its location in energy and absolute intensity. The key ingredient for the success of the NEA in this later case lies in its ability to capture non-Condon effects by sampling molecular geometries away from the equilibrium geometry (i.e., the transition dipole moments in Eq.~\eqref{eq:nea} depends on the nuclear configuration of the sampled geometry, $\boldsymbol{\mu}_{0I}(\mathbf{R}_j)$). Hence, the NEA was successfully used in the context of atmospheric chemistry to describe the low-energy $n\pi^\ast$ band of the photoabsorption cross-section for carbonyl-containing VOCs.\cite{hutton2022,marsili2022theoretical} Another potential limitation of the NEA is its inaccurate description of the tail of the photoabsorption cross-section, which should be stressed in the context of atmospheric photochemistry as a photolysis rate coefficient can be highly sensitive to a low-energy tail of the photoabsorption cross-section if it enters artificially deeper in the actinic region.

The NEA relies on the determination of an (approximate) ground-state nuclear probability density. The commonly employed approach is to use a harmonic approximation for the different vibrational modes of the molecule of interest, as one can determine analytically from this approximation a so-called Wigner distribution for both nuclear positions and momenta.\cite{persico2014overview,suchan2018importance,prlj2021calculating} The key advantage of this approach is that a harmonic Wigner distribution only requires, for each conformer, an equilibrium geometry in the ground electronic state and its corresponding vibrational frequencies -- two quantities readily available from any standard quantum-chemical software. Sampling molecular geometries from a harmonic Wigner distribution (used as a proxy for the ground-state nuclear probability density) is often called ‘Wigner sampling’. Wigner sampling is implemented in different codes to perform excited-state molecular dynamics, as the nuclear positions and momenta it provides can be used as initial conditions for excited-state dynamics. Using a harmonic Wigner distribution to provide the molecular geometries (nuclear positions) required by the NEA leads to accurate photoabsorption cross-sections for VOCs exhibiting rather harmonic modes (for example, acrolein\cite{prlj2021calculating}). However, care should be taken for flexible molecules, typically possessing low-frequency (anharmonic) normal modes, as they may lead to artifacts in the photoabsorption cross-section when the low-frequency mode is photoactive.\cite{prlj2023deciphering} More advanced strategies can be used to obtain ground-state distributions for flexible molecules, such as ground-state ab initio molecular dynamics with a quantum (or colored-noise) thermostat\cite{ceriotti2009nuclear,ceriotti2010colored,Finocchi2022} (QT) or path-integral molecular dynamics.\cite{markland2018nuclear} A detailed discussion of the NEA applied to VOC and its performances can be found in Ref.~\citenum{prlj2021calculating}.

From a practical perspective, calculating the photoabsorption cross-section for a given VOC requires the following steps. First, one needs to find all molecular conformers that are meaningfully present at a given temperature (this step can be done manually or involve dedicated software for conformer exploration). Once the conformers are located, an electronic-structure method should be chosen for the subsequent ground- and excited-state calculations, ensuring that we have the best compromise between computational efficiency and accuracy (this step requires a quantum-chemical package). Once the electronic-structure method is selected, the following steps should be executed for each conformer: (i) geometry optimization and frequency calculation, (ii) sampling of $N_p$ molecular geometries from the constructed harmonic Wigner distribution (this step often requires an external software), (iii) calculation of excitation energies and transition dipole moments for each sampled geometry, (iv) construction of the photoabsorption cross-section for the conformer (this step requires a bespoke code, sometimes the same as in step (ii)). The final photoabsorption cross-section is then calculated by weighting the photoabsorption cross-section of each conformer by its corresponding Boltzmann factor. The steps described above hopefully make it clear that determining the photoabsorption cross-section for a given VOC is not \textit{per se} difficult, but it implies a rather tedious and repetitive set of operations, involving the use of different software packages. In the next Sections, we discuss how these steps have been turned into an automated workflow within \atmo{}. 

\section{A guided tour of \atmo{}}
\label{sec:demonstration}

\begin{figure}[!ht]
    \centering
    \includegraphics[width=0.4\textwidth]{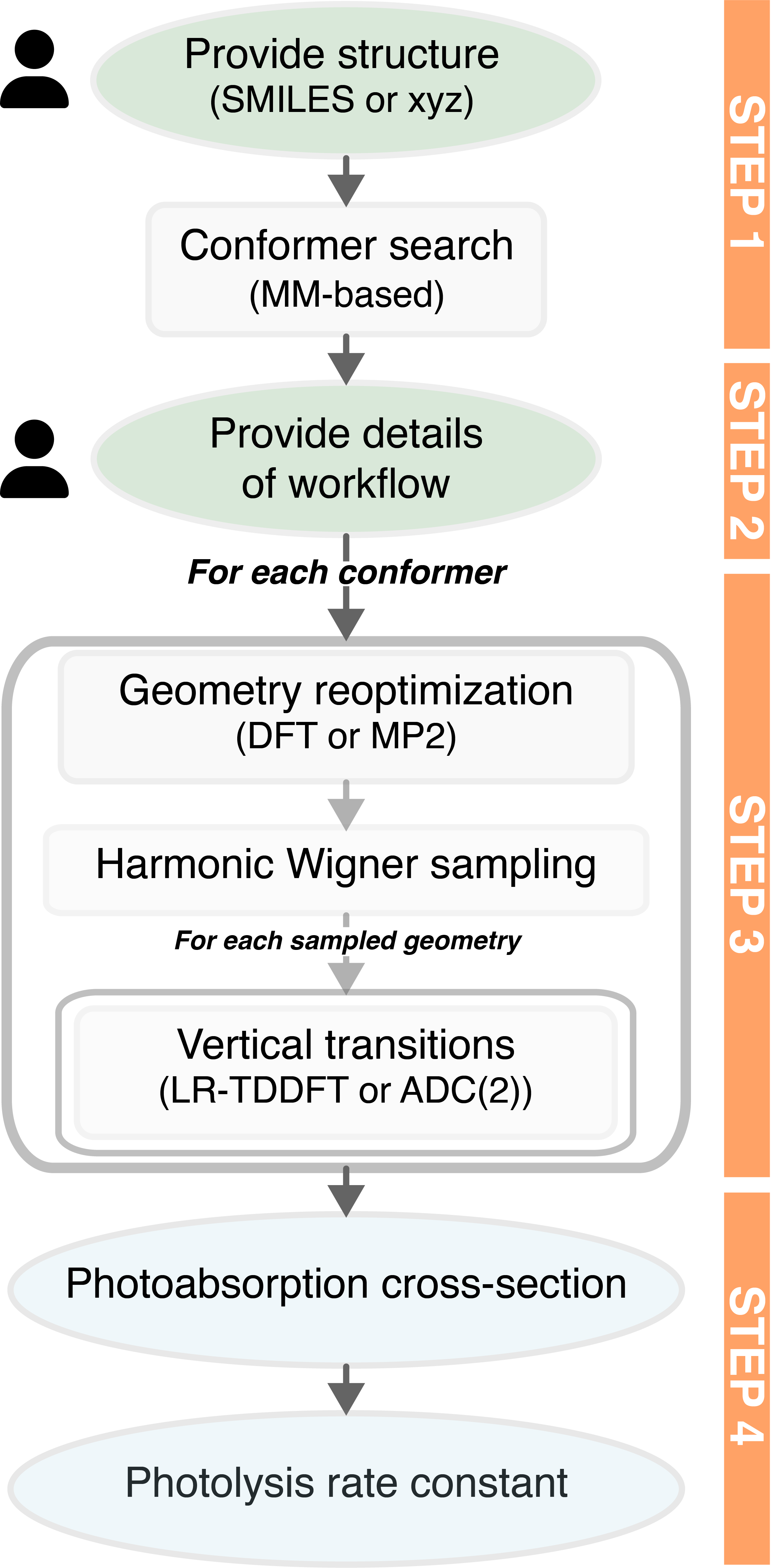}
    \caption{The different steps of \atmo{} and its overall workflow. User icons highlight the steps where the user input is requested. The step numbering corresponds to the subsections discussed in Sec.~\ref{sec:demonstration}. }
    \label{fig:workflow}
\end{figure}

The goal of \atmo{} is to transform the steps of an NEA calculation (described in the last paragraph of Sec.~\ref{sec:NEA}) into an automated workflow with the minimum effort for the user. The overall workflow of \atmo{} is depicted in Fig.~\ref{fig:workflow} and this Section endeavors to describe each step of an NEA calculation with \atmo{} from a user perspective. As an example, we propose to calculate the photoabsorption of glycolaldehyde, \ce{HCOCH2OH}.

\subsection{Step 1 -- Initial molecular structure and conformer search}

\begin{figure}[!ht]
    \centering
    \includegraphics[width=1.0\textwidth]{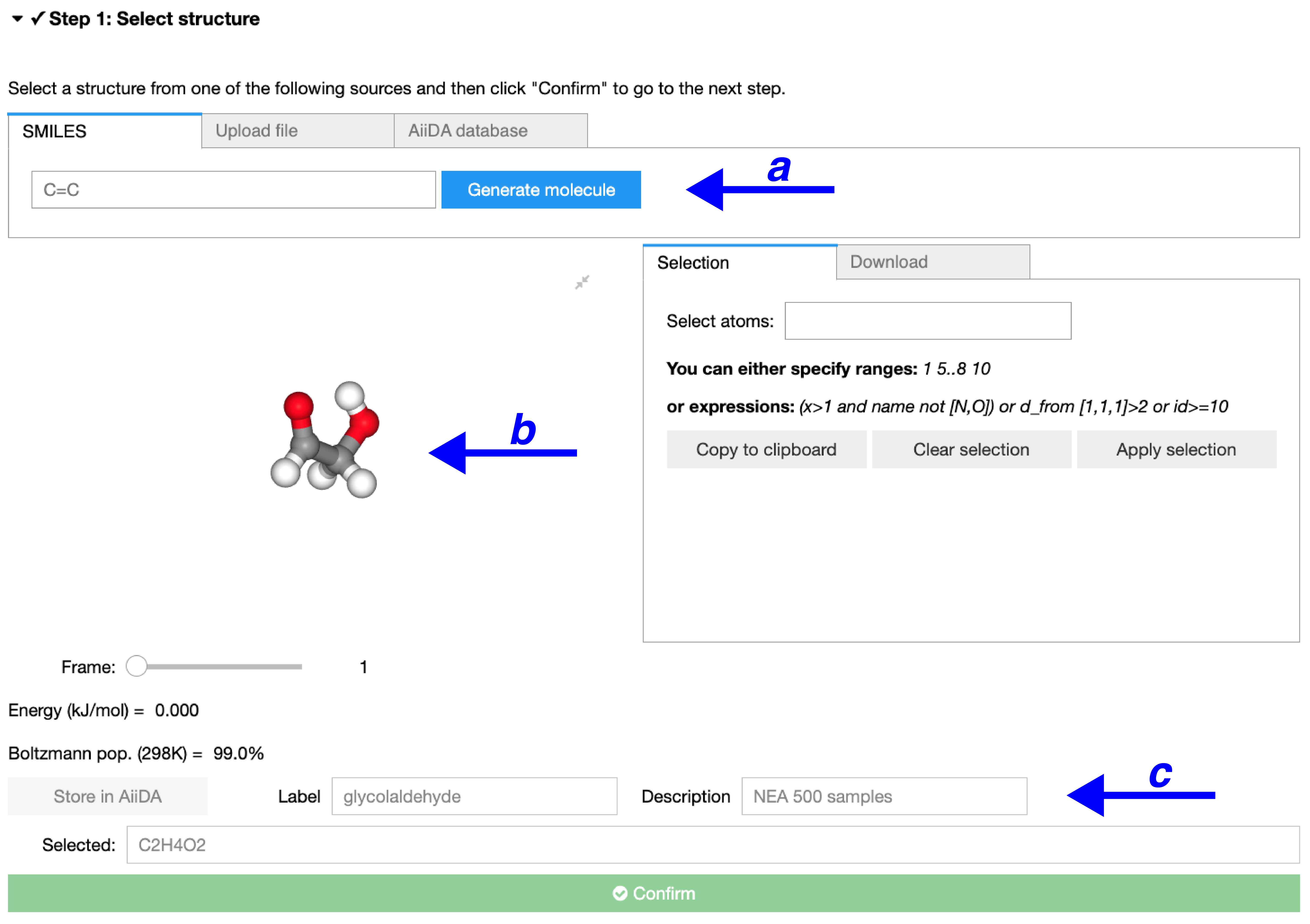}
    \caption{Step 1 -- (a) the user inputs the structure of the VOC of interest (SMILES code or an external coordinate file) and (b) \atmo{} proceeds with a (low-level) search of the dominant conformers, displaying all the conformers found and their approximate energy. (c) \atmo{} creates an entity in a database for this molecule.}
    \label{fig:atmospec1}
\end{figure}

\atmo{} implements its graphical interface via a web application that can be accessed using any modern web browser. The user connects to \atmo{} via its address on the machine where the code was installed (detailed installation information is provided on the GitHub page of \atmo{}\cite{daniel_hollas_2024_11075300})
After launching \atmo{}, the user is first requested to provide the molecular structure for the VOC of interest, here glycolaldehyde (Figure~\ref{fig:atmospec1}a). The simplest way to provide this information is to use a SMILES (simplified molecular-input line-entry system) code -- a simple ASCII string encoding the structure of a molecule.\cite{smilescode}
A SMILES code is easily obtained from a Lewis structure created with a molecular editor (for example, the SMILES code of glycolaldehyde is \texttt{O=CCO}). 
Once the SMILES code is provided, the user can click on the 'Generate molecule' button. 
\atmo{} will proceed with a fast search of the possible conformers of the molecule using the ETKDG algorithm\cite{rdkit:etkdg,rdkit:etkdgv3} implemented in the RDKit package,\cite{rdkit} followed by geometry optimization using the MMFF94 force field.\cite{MMFF94-IV} 
\atmo{} will display all the conformers of the molecule in a ball-and-stick representation within a window for the user to confirm that this is indeed the molecule of interest (Figure~\ref{fig:atmospec1}b). \atmo{} also displays the MMFF94 energies that serve as an initial estimate of their Boltzmann population. Both the conformer geometries and energetics are further refined in subsequent steps.
As an alternative to the input via SMILES, the user can also upload a file containing the molecular structure of interest, for example in an XYZ format. In this case, \atmo{} does not perform a conformer search.

The user is then invited to provide a description and label for the calculation to generate an entry in \atmo{} database (Figure~\ref{fig:atmospec1}c). The use of a database means that the results of this calculation can be recovered at any time in the future by selecting the entry via its label.

\subsection{Step 2 -- Computational details for the NEA calculation}

\begin{figure}[!ht]
    \centering
    \includegraphics[width=1.0\textwidth]{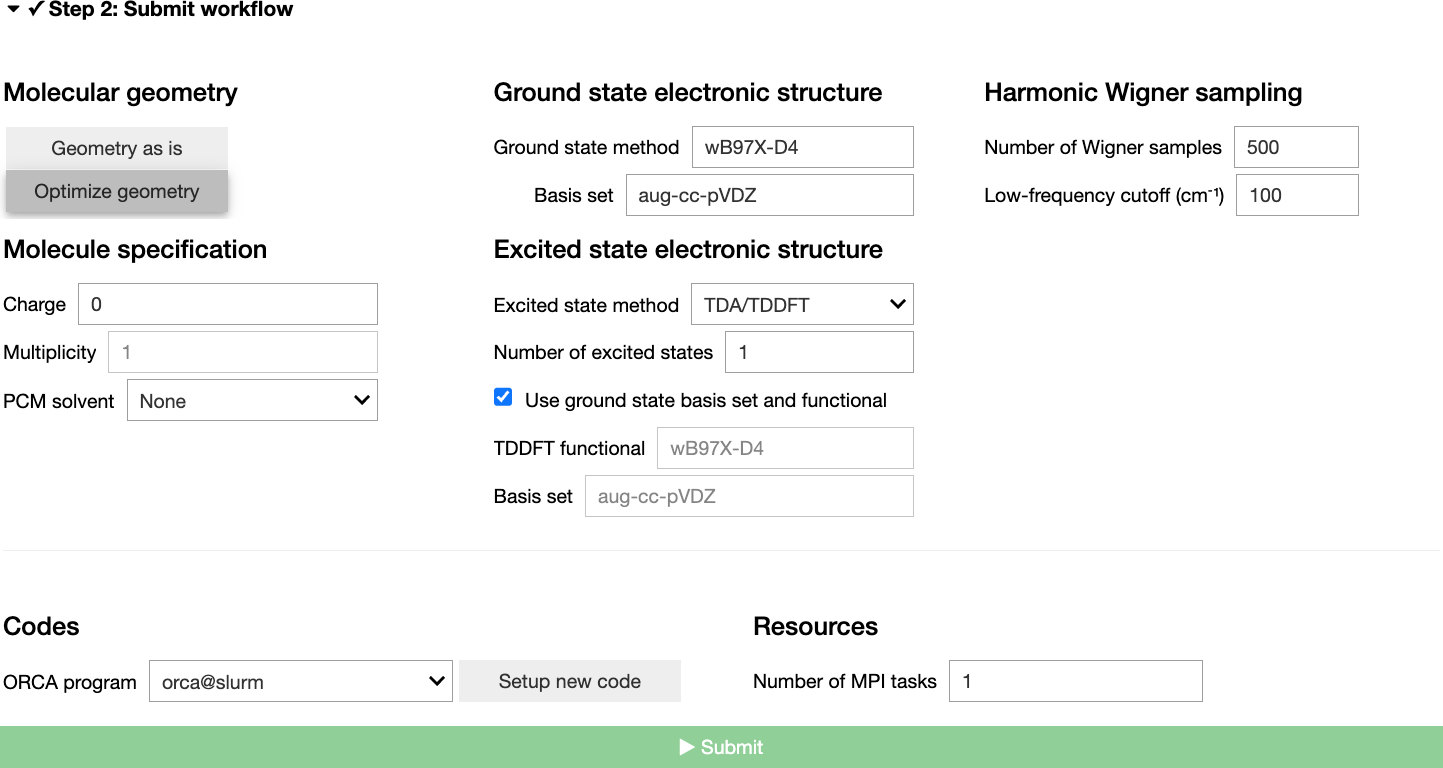}
    \caption{Step 2 -- The computational details of the \atmo{} workflow are provided. The user can specify the molecular details (charge and multiplicity) as well as the level of theory required. The number of sampled geometries from the harmonic Wigner distribution can be entered. Information about the computational resources (number of cores) can be defined. }
    \label{fig:atmospec2}
\end{figure}

The second step is critical as it defines the computational details for the calculation of the photoabsorption cross-section. The user will be invited to provide information about the molecule (charge, multiplicity) and the level of electronic-structure theory required. The ground-state electronic structure information will be used to optimize the different conformers identified in Step 1 and calculate the vibrational frequencies required to build the harmonic Wigner distribution for the molecule. The excited-state electronic structure method defines the level of theory for the calculation of vertical excitation energies and transition dipole moments that will be required for building the photoabsorption cross-section based on Eq.~\eqref{eq:nea}. In its current version, \atmo{} is interfaced with the quantum-chemical software package ORCA,\cite{orcarev} which offers a broad range of capabilities for excited electronic states and is free for academic use.

While default parameters are provided for the calculation of a photoabsorption cross-section with a compromise between efficiency and accuracy (based on earlier experience with the NEA and VOCs), \atmo{} offers different options to the users to refine their calculation and, perhaps more importantly, benchmark the level of electronic-structure theory. Hence, if the 'Number of Wigner samples' is set to 0, the user can perform single-point calculations on top of the optimized geometry of each conformer to test the sensitivity of vertical transitions to the quantum-chemical method. \atmo{} gives access to LR-TDDFT (with and without the Tamm-Dancoff approximation), but allows the user to perform calculations with the wavefunction-based methods ADC(2) and EOM-CCSD. While these two electronic-structure methods are likely to be quite expensive for a full NEA calculation (i.e., with multiple Wigner samples), they offer an adequate point of comparison for single-point calculations. We note that implicit solvent can be included too, motivated by recent experiments where the photoabsorption cross-section of VOCs was deduced from their extinction coefficients in cyclohexane.\cite{acp-23-7767-2023}

Once all the computational details are provided, the user can define where the calculations should be launched (locally or remotely, for example on an HPC cluster that the user has access to) as well as the number of cores that the electronic-structure software can use. By clicking on the 'Submit' button, the user will launch the fully-automatized NEA calculation.

\subsection{Step 3 -- Status of the calculation and outputs}

\begin{figure}[!ht]
    \centering
    \includegraphics[width=1.0\textwidth]{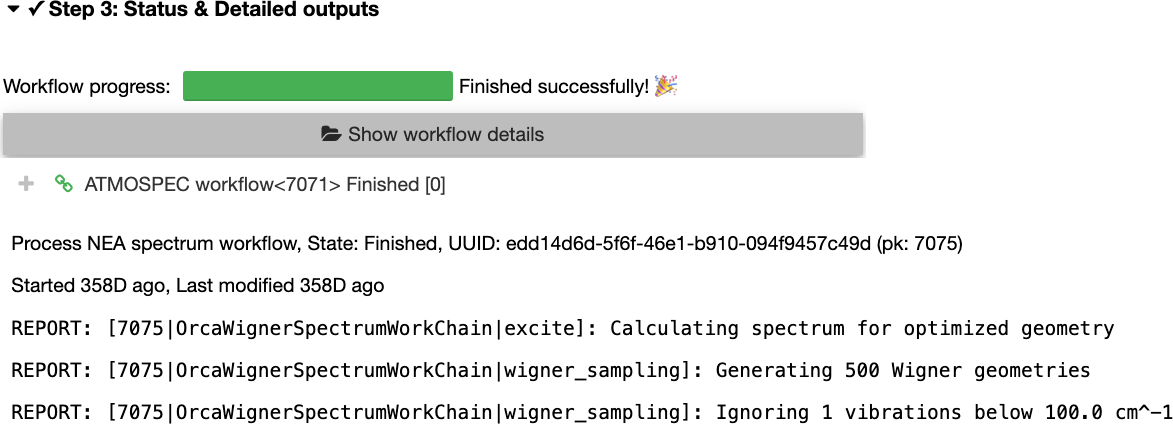}
    \caption{Step 3 -- This menu appears when an NEA calculation with \atmo{} has been launched and offers an overview of its progress. Clicking on 'Show workflow details' gives access to the underlying output files produced by the quantum-chemical software. All the output files are saved in the \atmo{} database.}
    \label{fig:atmospec3}
\end{figure}

Once the user clicks on the 'Submit' button in Step 2, \atmo{} distributes the required calculations on the available resources (either in a serial or parallel mode). The user sees a status bar appearing (Figure~\ref{fig:atmospec3}) that summarizes the progress of the NEA workflow. At this stage, the user can close the \atmo{} tab from their web browser, as the details of the molecule and its NEA workflow are saved in the \atmo{} database under the descriptor provided in Step 1. The user can simply reopen \atmo{} and search for the descriptor name in the \atmo{} database to reload the current status of the calculation. 

For the user interested in the details of the NEA workflow, Step 3 gives direct access to all the output files provided by the quantum-chemical software in use (in the present example, ORCA). The button 'Show workflow details' (Figure~\ref{fig:atmospec3}) opens the organization tree of the workflow -- geometry optimization and frequency calculation for each conformer, followed by the NEA calculation for each conformer -- and all the respective output files can be found at the bottom of this tree. At a higher level of this tree, \atmo{} provides a report of the key quantities extracted from the output files. 

\subsection{Step 4 -- Results}

\begin{figure}[!ht]
    \centering
    \includegraphics[width=1.0\textwidth]{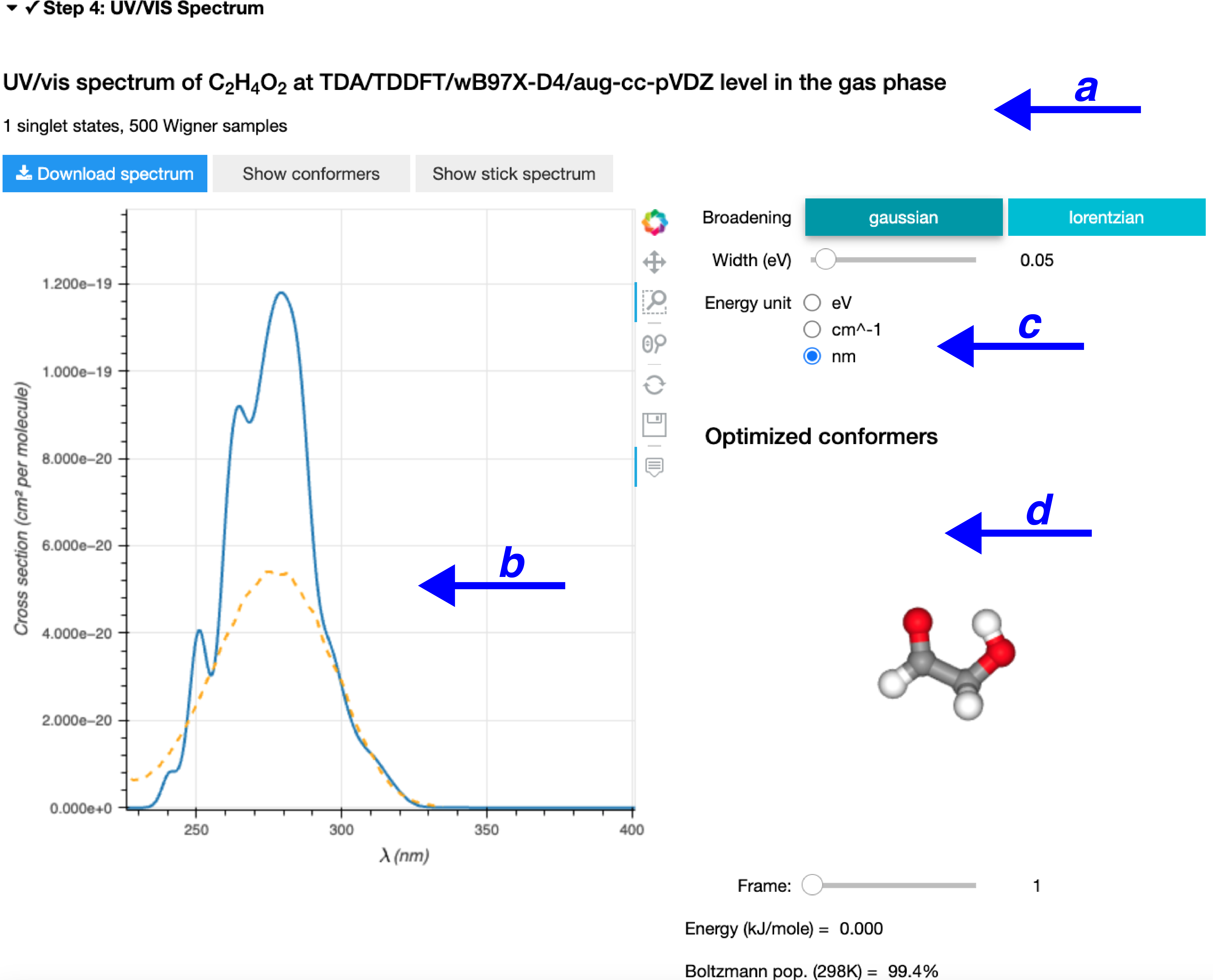}
    \caption{Photoabsorption cross-section produced by \atmo{}. (a) The top panel provides the details of the NEA workflow employed to produce the results presented in Step 4. (b) Interactive visualization of the calculated photoabsorption cross-section (blue) superimposed with the experimental cross-section (dashed orange) if uploaded by the user. (c) The representation of the photoabsorption cross-section (and its parameters) can be altered directly from the control panel. (d) The low-energy conformers of the molecule of interest are represented as ball-and-stick models in an interactive window, with their respective energy and Boltzmann population calculated at the level of theory defined in Step 2.}
    \label{fig:atmospec4}
\end{figure}

Once all the calculations are done, \atmo{} summarizes the results in the 'Step 4' and displays first the calculated photoabsorption cross-section (Figure~\ref{fig:atmospec4}). \atmo{} provides a summary of the computational details for the NEA workflow (Figure~\ref{fig:atmospec4}a). In this particular case, we calculated the photoabsorption cross-section of glycolaldehyde with LR-TDDFT/TDA, using the $\omega$B97X-D4 functional and the aug-cc-pVDZ basis set. 500 geometries were sampled from the harmonic Wigner distribution produced for each conformer (two conformers considered in total). 

The calculated photoabsorption cross-section is provided in an interactive window, together with any experimental photoabsorption cross-section that can be imported by the user (Figure~\ref{fig:atmospec4}b). In this particular example, the overall shape and width of the photoabsorption cross-section are properly captured by the NEA, while the intensity is slightly overestimated with the level of electronic-structure theory selected. Different tools are available to interact with the photoabsorption cross-section, and the user can save the results either as a CSV text file or a picture. A menu is also provided for tuning more advanced parameters related to the NEA (Figure~\ref{fig:atmospec4}c), such as the function used to broaden each vertical excitation within the NEA, its width ($\delta$), or the overall energy unit used to represent the photoabsorption cross-section. A side window depicts the low-energy conformers considered to build the photoabsorption cross-section (Figure~\ref{fig:atmospec4}d) and their relative energy, calculated at the level of theory defined in Step 2.

\begin{figure}[!ht]
    \centering
    \includegraphics[width=1.0\textwidth]{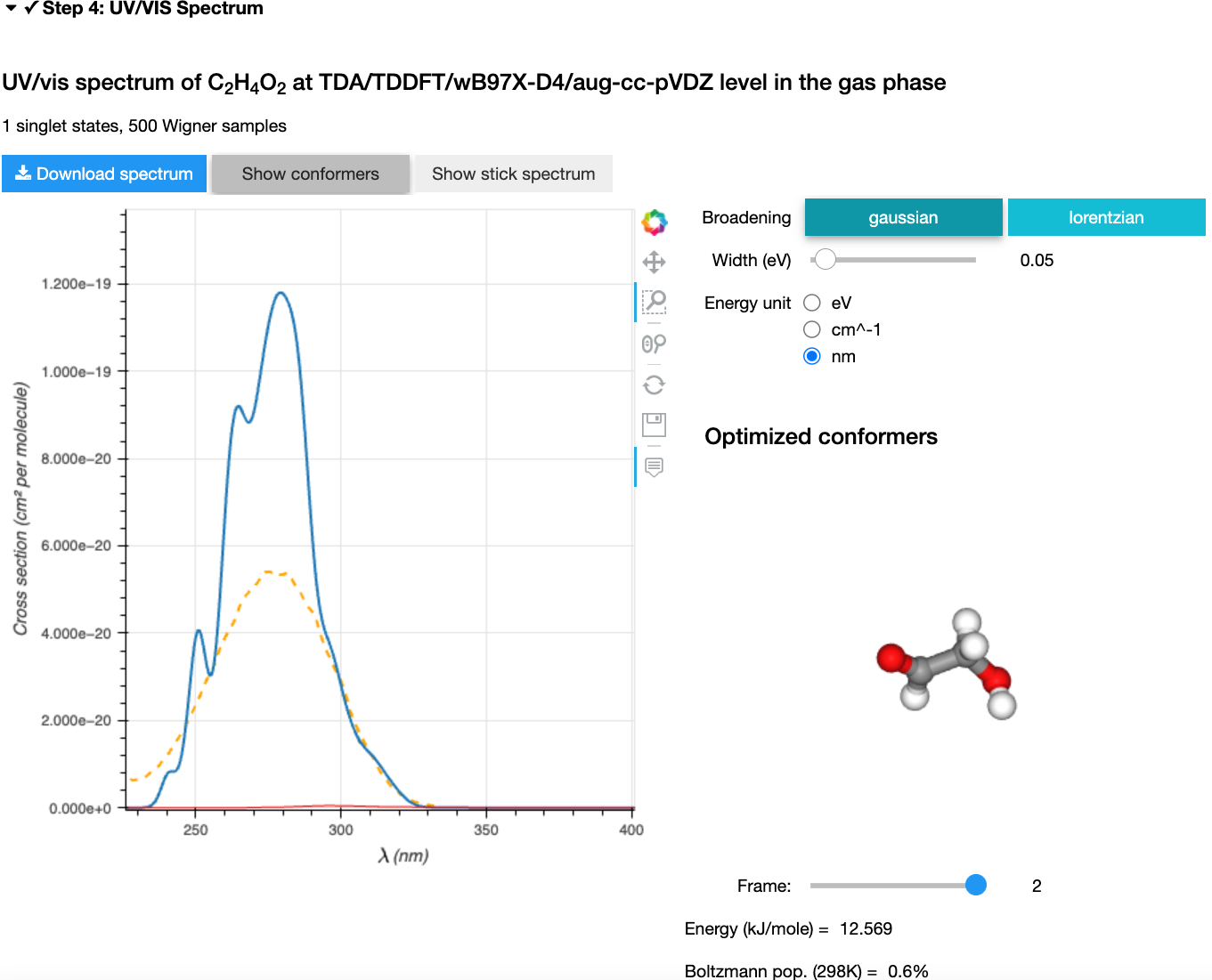}
    \caption{\atmo{} contains a tool to highlight the contribution of each conformer considered to the calculated photoabsorption cross-section. The red line in the photoabsorption cross-section window symbolizes the contribution to the photoabsorption cross-section by the selected conformer in the molecular window.}
    \label{fig:atmospec6}
\end{figure}

The analysis of the conformers can be pushed further by clicking on the 'Show conformers' button (Figure~\ref{fig:atmospec4}a). In this mode, \atmo{} shows the contribution of each conformer to the full photoabsorption cross-section. The user can select a conformer from the molecular-representation window, and its photoabsorption cross-section -- weighted by the Boltzmann population of the conformer -- is highlighted in red in the photoabsorption cross-section (Figure~\ref{fig:atmospec6}). In the specific case of glycolaldehyde presented here, a low-energy conformer with a Boltzmann population of 0.6\% at 298 K shows a contribution to the photoabsorption cross-section that is shifted to longer wavelength (red line in Figure~\ref{fig:atmospec6}) in comparison to the dominant conformer exhibiting an intramolecular hydrogen bond (blue line in Figure~\ref{fig:atmospec6}). 

\begin{figure}[!ht]
    \centering
    \includegraphics[width=0.8\textwidth]{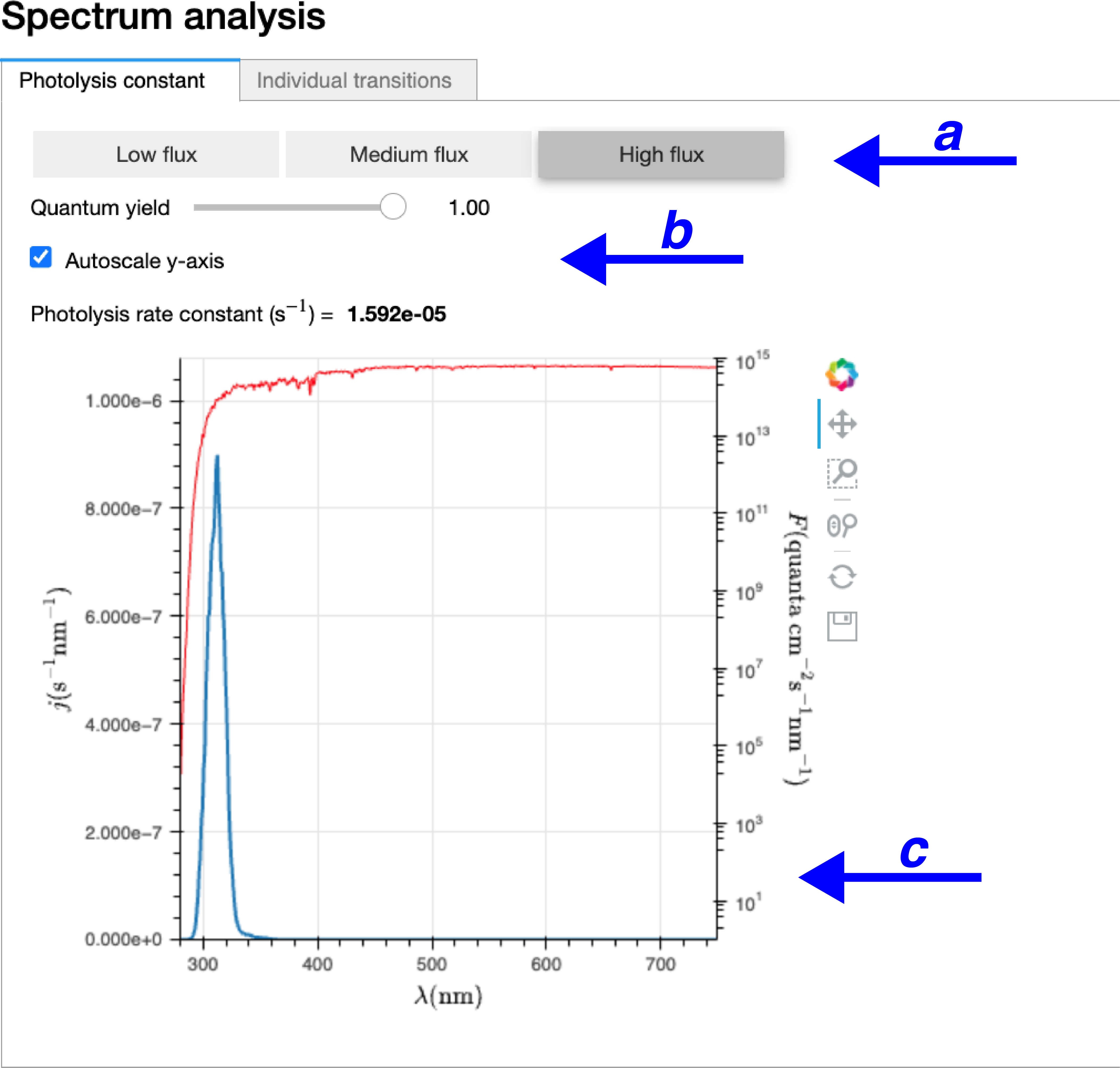}
    \caption{Qualitative prediction of the photolysis rate constant for the VOC of interest based on the calculated photoabsorption cross-section. (a) The user can select three standardized actinic fluxes $F(\lambda)$ -- low, medium, or high flux (see main text for details). (b) The quantum yield is set by the user and allows them to determine a qualitative value for the photolysis rate constant by integrating Eq.~\eqref{eq:photolysis} over the actinic flux. (c) Interactive window showing the integrand  $j(\lambda)=\phi(\lambda)\sigma(\lambda) F(\lambda)$ (blue curve) and the actinic flux (red curve).}
    \label{fig:atmospec7}
\end{figure}

A last window gives access to an estimation of the photolysis rate constant $J$ using Eq.~\eqref{eq:photolysis} with the calculated photolysis rate constant, a user-defined value for the photolysis quantum yield, and three options for the actinic fluxes (Figure~\ref{fig:atmospec7}). The user can select one of the three standardized actinic fluxes (Figure~\ref{fig:atmospec7}a) -- 'high flux' (solar zenith angle = 0$^\circ$, overhead ozone column = 200 DU), 'medium flux' (solar zenith angle = 60$^\circ$, overhead ozone column = 350 DU), and 'low flux' (solar zenith angle = 90$^\circ$, overhead ozone column = 500 DU) for a ground elevation of 0 km above sea level. An interactive window depicts the integrand of Eq.~\eqref{eq:photolysis}, namely the product of the calculated photoabsorption cross-section, the quantum yield, and the chosen actinic flux (Figure~\ref{fig:atmospec7}c). An approximate \textit{wavelength-independent} quantum yield can be provided by the user (Figure~\ref{fig:atmospec7}b) and its effect on the calculated $J$ value is adapted in real-time so that the user can determine the upper and lower bounds for the photolysis rate constant based on the photoabsorption cross-section predicted by \atmo{}.

\section{Software architecture and implementation}
\label{sec:architecture}

In this Section, we provide a brief description of the overall architecture of \atmo{} as well as an overview of the software stack and libraries used to build this tool.

\subsection{Design goals}
The overarching goal of \atmo{} was to make the calculation of photoabsorption cross-section with the NEA accessible to non-experts in computational photochemistry. With that in mind, the following criteria were key for the development of the architecture and implementation of \atmo{}: (a) simple installation, (b) graphical interface that is easy to navigate, (c) input and output data should be stored, and easily searchable, and (d) the calculations should be fully reproducible.

A trivial way to address point (a) is to develop a software that does not necessarily need to be installed on the user's computer. This can be achieved by making the graphical interface run in a web browser, while the application itself may run on a remote computer. This strategy means that the application itself can be deployed by an expert user or IT service, while the user only needs to navigate to the correct URL in the web browser to interact with its graphical interface. As a secondary goal, we also aimed to make the installation of the application approachable and reproducible. We achieved that by delivering \atmo{} as a Linux container (using Docker\cite{merkel2014docker}), which bundles all the necessary dependencies together.
To achieve points (c) and (d), we utilized the workflow manager \aiida{} (see below). 
We note that the points highlighted above align with the FAIR principles (findable, accessible, interoperable, reusable) for scientific data.\cite{Wilkinson:2016aa} To further enhance the reusability of \atmo{}, the different parts of the software (such as the harmonic Wigner sampling module) will be made available as standalone Python packages, usable independently from the full \atmo{} tool.

\subsection{Architecture and data flow}

\begin{figure}[!ht]
    \centering
    \includegraphics[width=1.0\textwidth]{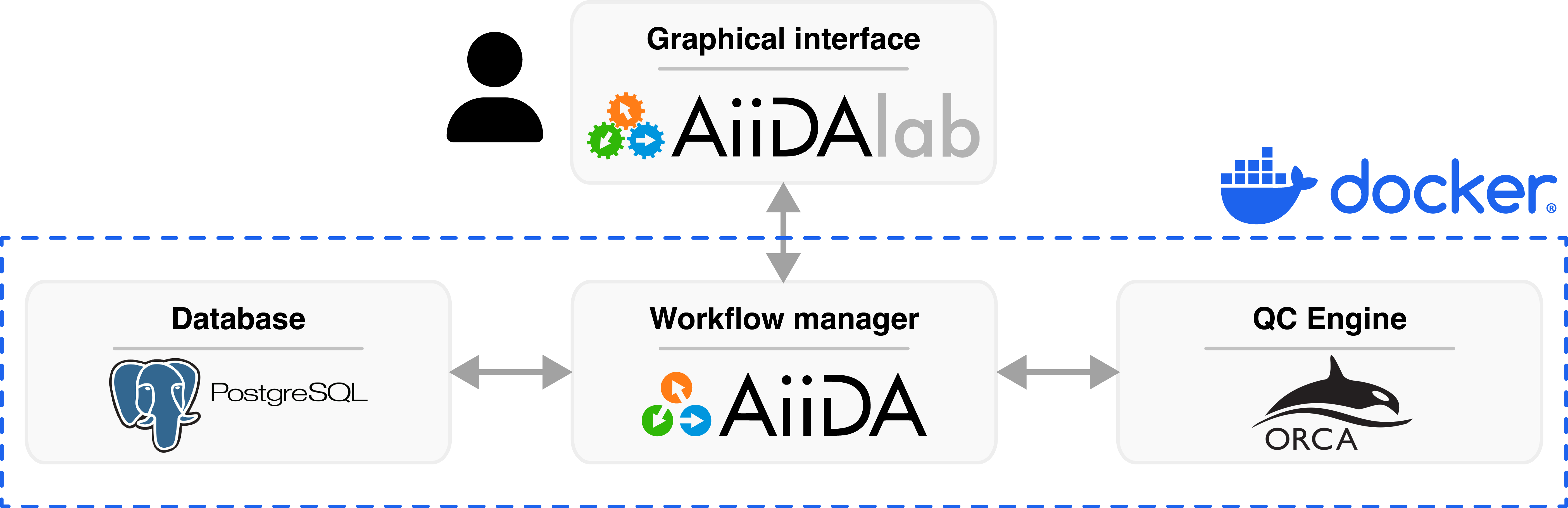}
    \caption{Architecture diagram of \atmo{}. The Graphical interface in the browser (implemented in the \aiidalab{} framework) communicates with the Workflow manager (\aiida{}), which orchestrates the \textit{ab initio} calculations (using ORCA) and stores the results in the SQL Database (PostgreSQL). Each rectangle in the diagram may represent a different physical computer, but the whole system can be deployed on a single computational server. The individual components can be deployed together using a Docker container.}
    \label{fig:architecture}
\end{figure}

The overall architecture of \atmo{} is depicted in Fig.~\ref{fig:architecture}. We have decided to utilize the existing ecosystem around the workflow manager \aiida.\cite{UHRIN2021110086,Huber:2020aa} With this choice, the implementation of the NEA workflow is fairly generic: \aiida{} handles the tasks of submitting individual \textit{ab initio} calculations to the quantum-chemical program (here, ORCA), parsing the resulting outputs and storing all the critical information in a relational database (PostgreSQL) and a custom file storage. This strategy gives us the flexibility to combine \atmo{} with other quantum-chemical programs in the future. Moreover, \aiida{} benefits from broad support for handling automatic job submission to remote HPC clusters (key to install \atmo{} locally but utilize a remote HPC resource for the quantum-chemical calculations).

The other central part of \atmo{} is its graphical interface (front-end). We again employed \aiidalab, which was specifically created to make quantum-chemical calculations and complex scientific workflows available through a user-friendly browser interface.\cite{aiidalab2021} In \aiidalab, the front-end is built on top of a Jupyter notebook interface. While this choice is somewhat unconventional for a web application, it has the advantage that the graphical part of the code is written in the same programming language (Python) as the rest of the program. This strategy makes \atmo{} (and other \aiidalab-based applications) more approachable for potential developers, in a scientific environment where Python is nowadays perceived as a \textit{de facto lingua franca}.

The AtmoSpec code is fully open source, as is most of the software stack that it builds upon, with the major exception of ORCA, which is however available for free for academic use under a custom EULA license. (The link to the \atmo{} GitHub repository can be found in Ref.~\citenum{daniel_hollas_2024_11075300}.)

\section{Caveats and future developments}
\label{sec:caveats}

\atmo{} provides a simple interface to calculate an approximate photoabsorption cross-section, but the user needs to keep in mind that this simplicity leads to a series of strict \textit{limitations}. We list here the main current limitations, and we discuss in the following paragraph the future developments that may address some of these issues. (1) \atmo{} only offers a rather modest choice of (single-reference) electronic-structure methods and, as such, is not equipped to treat molecules exhibiting a challenging electronic structure (e.g., radicals or metal complexes). (2) \atmo{} relies on the NEA, meaning that the produced photoabsorption cross-sections do not exhibit any vibronic structure. Perhaps more importantly in the atmospheric context, the (low-energy) tail of the photoabsorption cross-sections is likely to be too high in intensity, even for purely dissociative states. (3) \atmo{} uses a harmonic Wigner distribution to sample the geometries used in the NEA. This strategy may lead to artifacts when used for flexible VOCs, that is, molecules exhibiting low-frequency vibrational modes (< 200 cm$^{-1}$), as discussed in Ref.~\citenum{prlj2021calculating}. Such modes can be removed from the sampling by applying a low-frequency cutoff in Step 2. The use of \textit{ab initio} molecular dynamics combined with a quantum thermostat is desirable for sampling the ground-state probability density of such molecules, but harder to automatize. (4) \atmo{} does not offer any speed-up for the calculation of photoabsorption cross-sections within the NEA. For VOCs, these calculations can take from a few hours to a few days, depending on the level of theory and the number of sampled geometries for the NEA. 

A series of future developments are planned for \atmo{}, some of them trying to address the limitations listed above. (a) To reduce the computational cost of a typical photoabsorption cross-section calculation, we plan to interface \atmo{} with the GPU-accelerated electronic structure package TeraChem. In addition to this speed-up of the electronic-structure calculations, we will incorporate the idea of optimal sampling within \atmo{}.\cite{optimalsampling2021} The idea of the optimal sampling is to use a cheap level of electronic-structure theory (for example, ZIndo/S) to calculate the photoabsorption cross-section of the molecule of interest within the NEA and a normal number of sampled geometries (around 500). From this result, the optimal sampling can select a subset (around 50) of optimal geometries that best represent the overall photoabsorption cross-sections. The final photoabsorption cross-section can be obtained by performing the NEA on this subset of geometries using a higher level of electronic structure theory. (b) As part of the automation in \atmo{}, we plan to devise a strategy to pre-select an optimal level of theory based on the input molecular structure and automatically determine the number of excited electronic states important to describe the absorption of the molecule in the actinic region. (c) \atmo{} could be generalized to other types of spectroscopies within the NEA framework, such as photoelectron spectroscopy\cite{doi:10.1021/acs.jctc.6b00704,clarke2022photochemistry} or X-ray absorption spectroscopy.\cite{D2CP05299G}

\section{Summary}

In summary, this work introduced \atmo{}, a computational tool devised to predict the photoabsorption cross-section and other photolysis properties of transient volatile organic compounds for atmospheric applications. \atmo{} offers an automated workflow for the nuclear ensemble approach, only requiring a minimal input from the (non-expert) user using a web browser interface. This work describes a typical calculation with \atmo{}, going through the details of each step of the workflow. We also highlighted the main limitations of the current implementation and its future developments. \atmo{} is open source and freely available for download -- the GitHub link can be found in Ref.~\citenum{daniel_hollas_2024_11075300}. 

\begin{acknowledgement}

The authors would like to thank Fay Abu Al Timen, Marco Barnfield, Kirstin Gerrand, Will Hobson, Kon Nomerotski, and Emily Wright -- undergraduate students at the University of Bristol -- for their contribution to the development of the photolysis rate constant calculation widget. DH would like to thank the \aiida{} and \aiidalab{} developers that comprise an incredibly welcoming open source community, in particular Aliaksander Yakutovich, Jusong Yu, Xing Wang, Marnik Bercx, Sebastiaan Huber, Carl Simon Adorf, Carlo Pignedoli, and Giovanni Pizzi. DH would also like to thank the many users of AtmoSpec within the group in Bristol for their feedback and bug reports. The authors also acknowledge Sasha Madronich for providing the standardized actinic fluxes implemented in \atmo{}.  This project has received funding from the European Research Council (ERC) under the European Union's Horizon 2020 research and innovation programme (Grant agreement No. 803718, project SINDAM) and the EPSRC Grants EP/V026690/1, EP/X026973/1, and EP/Y01930X/1. This article is also based upon work from COST Action CA18212 - Molecular Dynamics in the GAS phase (MD-GAS), supported by COST (European Cooperation in Science and Technology).

\end{acknowledgement}


\providecommand{\latin}[1]{#1}
\makeatletter
\providecommand{\doi}
  {\begingroup\let\do\@makeother\dospecials
  \catcode`\{=1 \catcode`\}=2 \doi@aux}
\providecommand{\doi@aux}[1]{\endgroup\texttt{#1}}
\makeatother
\providecommand*\mcitethebibliography{\thebibliography}
\csname @ifundefined\endcsname{endmcitethebibliography}
  {\let\endmcitethebibliography\endthebibliography}{}

\end{document}